\documentclass[aps,prl,twocolumn,showpacs]{revtex4}
\usepackage{graphicx} 
\usepackage{dcolumn}  
\usepackage{bm}       
\usepackage{amsmath}
\usepackage{epsfig}

\def\ii{{\rm i}}        

\begin{document}
\title{Robust plasmon waveguides in strongly-interacting nanowire arrays}
\author{A.~Manjavacas}
\affiliation{Instituto de \'Optica - CSIC, Serrano 121, 28006
Madrid, Spain}

\author{F.~J.~Garc\'{\i}a de Abajo}
\email[Corresponding author: ]{jga@cfmac.csic.es}
\affiliation{Instituto de \'Optica - CSIC, Serrano 121, 28006
Madrid, Spain}

\date{\today}

\begin{abstract}
Arrays of parallel metallic nanowires are shown to provide a
tunable, robust, and versatile platform for plasmon interconnects,
including high-curvature turns with minimum signal loss. The
proposed guiding mechanism relies on gap plasmons existing in the
region between adjacent nanowires of dimers and multi-wire arrays.
We focus on square and circular silver nanowires in silica, for
which excellent agreement between both boundary element method and
multiple multipolar expansion calculations is obtained. Our work
provides the tools for designing plasmon-based interconnects and
achieving high degree of integration with minimum cross talk between
adjacent plasmon guides.
\end{abstract}
 \maketitle


Electromagnetic modes in metal surfaces known as plasmons can
propagate along millimeters in metallic structures at near-infrared
frequencies \cite{S1981}, thus providing a plausible substitute for
the electrical impulses used in current electronic circuits
operating at microwave clock frequencies \cite{O06,ZSC06}. Several
designs of plasmon interconnects have been prototyped in recent
years, including metallic waveguides of finite cross section in
symmetric \cite{TYT97,B00} and asymmetric \cite{B01} environments,
channels cut into flat surfaces \cite{BVD06}, plasmon-band-gap
structures based upon periodic corrugations \cite{BEL01}, and
plasmon hopping in arrays of nanoparticles \cite{KDW99,MKA03}.
Plasmon modes can be tuned in frequency, and their spatial
distribution molded, by tailoring the geometry of metallic
structures on the nanometer scale.
In particular, extreme plasmon confinement has been achieved in
narrow insulator films buried inside metal \cite{MK06}. Actually,
buried structures provide a natural but technologically challenging
approach to compact integration. In contrast to that, open plasmonic
geometries involve electromagnetic fields extending significantly
away from the metal \cite{B00,B01,BVD06,BEL01,KDW99,MKA03}, and
consequently producing a substantial degree of cross-talk between
neighboring waveguides \cite{CSS07}.

In this Letter, arrays of parallel metallic nanowires are shown to
provide a versatile and tunable platform for highly-integrated
plasmon interconnects. The propagation distance and degree of
confinement of the plasmon guided modes depend strongly on the
separation between wires. Individual wire modes are recovered at
large separations, while mode hybridization is observed when the
spacing is reduced. Gap modes are observed at small separations,
highly localized in the regions between two adjacent wires.
We use both the boundary element method (BEM) \cite{Manja1} and a
two-dimensional multiple-elastic-scattering multipolar expansion of
the fields for straight cylinders (2D-MESME) \cite{Manja2}, with the
two approaches resulting in complete agreement on the scale of the
plots. These methods provide rigorous solutions of Maxwell's
equations in frequency space for materials described by local
dielectric functions and separated by abrupt interfaces. We focus on
silver nanowires of circular and square cross sections embedded in
silica. The dielectric functions of silver \cite{JC1972} and silica
\cite{P1985} have been taken from optical data. The proposed guiding
mechanism is demonstrated to be tolerant to asymmetry in wire dimers
and sharp turns of subwavelength radius.

The localized plasmons sustained by our structures can be
conveniently characterized using the photonic local density of
states (LDOS), defined by analogy to its electronic counterpart as
the combined local intensity of all eigenmodes of the system under
investigation. We in particular consider the LDOS relative to its
value in vacuum, $\omega^2/3\pi^2c^3$. The LDOS is proportional to
the radiative decay rate of excited atoms \cite{FMM04}, which we in
turn obtain using BEM from the imaginary part of the self-induced
electric field acting back on a dipole \cite{paper085}. We have
double checked our results by comparing with the excess of
space-integrated total density of states (DOS) with respect to
vacuum, which is directly accessible through 2D-MESME \cite{FMM04}.


\begin{figure}[ht]
\includegraphics[width=65mm,angle=0]{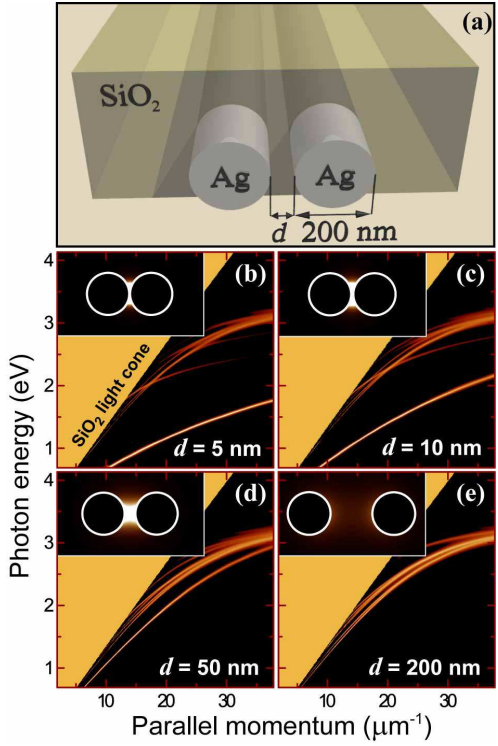}
\caption{Gap plasmon modes of two parallel silver nanowires in
silica. {\bf (a)} Schematic view of the geometry. {\bf (b)-(e)}
Photonic density of states (DOS) as a function of energy and
momentum parallel to the wires for various dimer separations $d$.
The insets show the spatial distribution of the local density of
states (LDOS) for the lowest-energy gap mode at a free-space light
wavelength of 1550\,nm. Brighter regions correspond to higher DOS
and LDOS. The maximum LDOS in the inset of (c) is $\sim 15000$ times
the vacuum value.} \label{Fig1}
\end{figure}

\begin{figure}[ht]
\includegraphics[width=65mm,angle=0]{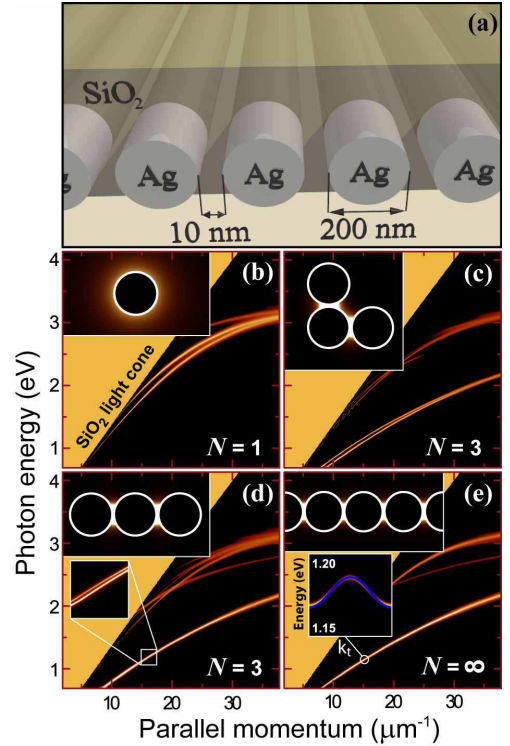}
\caption{Evolution of gap plasmon modes with the number of nanowires
in an array ($N$). {\bf (a)} Schematic view of the geometry. {\bf
(b)-(e)} DOS as a function of energy and momentum $k_\parallel$
parallel to the wires for arrays of $N=1,3,$ and $\infty$ nanowires.
The insets show LDOS maps for the lowest-frequency gap mode at a
wavelength of 1550\,nm. The lower inset in (e) shows the transverse
momentum dependence of the DOS for $k_\parallel=16\,\mu$m$^{-1}$
compared to a tight-binding model of the gap mode (solid curve) over
the first Brillouin zone of the 1D lattice.} \label{Fig2}
\end{figure}

We start by considering a dimer formed by two 200-nm circular silver
wires embedded in silica, as shown in Fig.\ \ref{Fig1}(a). The
contour plots of Fig.\ \ref{Fig1}(b)-(e) show the DOS resolved in
contributions of different momentum $k_\parallel$ parallel to the
wires for various separations between wire surfaces ($d$). A
strongly bound mode is observed at small wire separations [Fig.\
\ref{Fig1}(b)], with $k_\parallel$ well above $k_{\rm h}$, the
momentum of light in the host silica. The spatial extension of this
gap mode is limited to the inter-wire region (see inset), and thus,
it is expected to interact very weakly with other structures sitting
in the vicinity of the wires but far from the gap. This mode evolves
continuously for increasing inter-wire distance to become a
hybridized monopole-monopole mode of induced-charge pattern
$(+)\cdot\cdot\cdot(-)$ aligned with the dimer axis [Fig.\
\ref{Fig1}(c)-(d)]. This is in contrast to the
$(+-)\cdot\cdot\cdot(+-)$ dipole-dipole plasmon in particle dimers
\cite{paper114}, which is the lowest-energy mode according to
plasmon chemistry arguments \cite{NOP04}. In this sense, wires are
distinctly different from particles because charge neutrality is
guaranteed by oscillations along the rods for finite $k_\parallel$,
thus making two-dimensional monopoles possible. At sufficiently
large distance, single-wire plasmons of $m=0$ azimuthal symmetry are
recovered [cf. Fig.\ \ref{Fig1}(e) and Fig.\ \ref{Fig2}(b); see
Ref.\ \cite{AE1974} for analytical expressions of single-wire
plasmons].


Fig.\ \ref{Fig2} proves that the gap mode is really local. The
lowest-frequency plasmon branch of the trimers in Figs.\
\ref{Fig2}(c) and \ref{Fig2}(d) follows approximately the dispersion
relation of the gap mode in the dimer with the same gap distance
$d=10$\,nm [Fig.\ \ref{Fig1}(c)], although closer examination
reveals the splitting of this mode into two very close modes [see
lower inset in Fig.\ \ref{Fig2}(d)]. In the infinite wire array of
Fig.\ \ref{Fig2}(e) a plasmon branch of gap modes is found for each
value of the transverse momentum $k_t$, parallel to the array plane
and perpendicular to the wires ($k_t=0$ in the figure).
Incidentally, propagation across wires mimics plasmon hopping in
particle chains \cite{WL05}.

\begin{figure}[ht]
\includegraphics[width=60mm,angle=0]{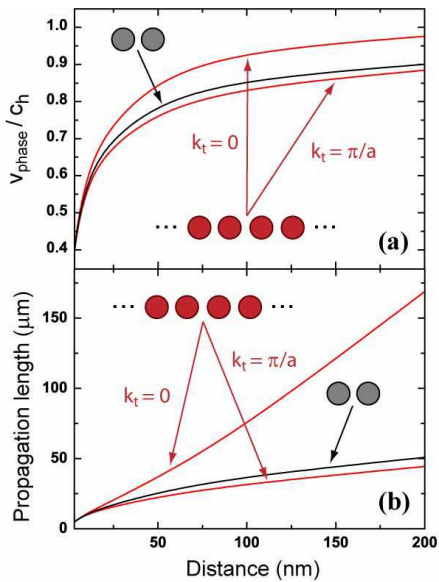}
\caption{{\bf (a)} Phase velocity of gap plasmon modes in silver
wire dimers and infinite-arrays (for $k_t=0$ and $k_t=\pi/a$) as a
function of separation $d$ for fixed wavelength $\lambda=1550$ nm.
{\bf (b)} Propagation length $L$ under the same conditions as in
(a), obtained from $L=1/2{\rm Im}\{k_\parallel\}$, where ${\rm
Im}\{k_\parallel\}$ corresponds to the HWHM of the
$k_\parallel$-dependent DOS.} \label{Fig3}
\end{figure}


Nearly-touching trimers can be regarded as dimers formed by two
coupled gaps, and similarly, $N$-wire arrays behave as structures
formed by $N-1$ gaps. The large degree of plasmon localization
observed for small $d$ suggests using a tight-binding model
\cite{AM1976}, in which an unperturbed Hamiltonian $H_0$ describes
uncoupled gap modes $|j\rangle$ at sites $j$, such that $\langle
j|H_0|j' \rangle=\omega_{k_\parallel}\delta_{jj'}$, where
$\omega_{k_\parallel}$ is the mode energy for fixed $k_\parallel$.
Neighboring gaps can interact in this model via a potential $V$,
with non-zero matrix elements $\langle j|V|j\pm
1\rangle=\Delta_{k_\parallel}/2$, where $\Delta_{k_\parallel}$ is
the hopping energy. Then, the two linear-trimer modes have energies
$\omega_{k_\parallel}\pm\Delta_{k_\parallel}/2$ [i.e., the dimer
band lies halfway between the two trimer bands, which we have
verified by comparison of Figs.\ \ref{Fig1}(c) and \ref{Fig2}(d)].
Also, the plasmon modes of an infinite, periodic wire array must
have the form $|\psi_{k_t} \rangle=\sum_j\exp(\ii k_taj)\,|j
\rangle$ in virtue of Bloch's theorem \cite{AM1976}, where $a$ is
the period. These states diagonalize the full Hamiltonian $H_0+V$
and have energies $\omega_{k_\parallel
k_t}=\omega_{k_\parallel}+\Delta_{k_\parallel}\cos(k_ta)$. We have
tested this formula in the array of Fig.\ 2(e), where the lower
inset shows $\omega_{k_\parallel k_t}$ as a function of $k_t$ for
$k_\parallel=16\,\mu$m$^{-1}$ (solid curve) compared with the actual
2D-MESME calculation of the DOS. The unperturbed gap energy
$\omega_{k_\parallel}=1.18$\,eV and the hopping parameter
$\Delta_{k_\parallel}=0.01$\,eV have been taken from the dimer
[Figs.\ \ref{Fig1}(c)] and trimer [Figs.\ \ref{Fig2}(d)] with the
same value of $d$, respectively. Similar agreement between model and
full calculation is observed over the range of $k_\parallel$ under
consideration.

Interestingly, these analytical expressions apply to the large $d$
limit as well, in which the tight-binding model is constructed based
upon localized plasmons of the wires, showing similar agreement with
rigorous DOS calculations. Tight-binding is thus the natural
description of both the small and large $d$ limits in the noted
continuous evolution from the localized gap mode (small $d$) to the
lowest-energy hybridized (monopole-monopole) wire modes (large $d$)
\cite{paper114}.

The degree of plasmon localization increases with decreasing gap
distance $d$. This is reflected in a reduction of the phase velocity
$v_p=c_hk_h/k_\parallel$ relative to the speed of light in silica,
$c_h$, as shown in Fig.\ \ref{Fig3}(a). The group velocity (no
shown) is also reduced, so that gap modes become considerably slower
than light in silica. The propagation distance is strongly-dependent
on inter-wire distance [Fig.\ \ref{Fig3}(b)]: the large confinement
observed at small separations increases the relative weight of the
electric field intensity inside the metal, where ohmic losses are
produced in proportion to that intensity within linear response.
Nevertheless, the gap mode involves electric field polarization
mainly perpendicular to the wire surfaces near the gap, where light
energy is concentrated, and this is beneficial to obtain longer
propagation distances because the normal electric field inside the
metal is reduced by its large dielectric function to fulfill the
continuity of the normal electric displacement. This gives rise to
propagation distances of the order of tens of microns for
separations of tens of nanometers, accompanied by relatively large
mode confinement. The tradeoff between confinement and propagation
distance is clearly illustrated in the long-$d$ behavior of the
infinite array for $k_t=0$ and $k_t=\pi/a$, with the former showing
longer propagation and smaller phase velocity (see Fig.\
\ref{Fig3}).

\begin{figure}[ht]
\includegraphics[width=65mm,angle=0]{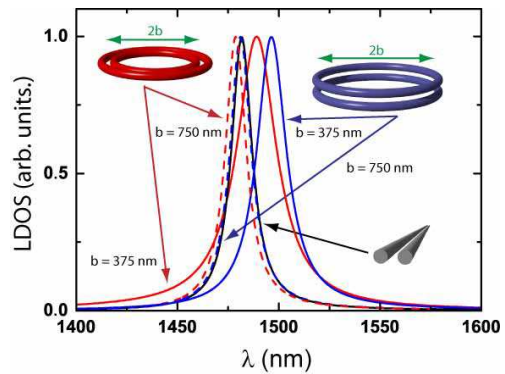}
\caption{Gap mode in co-planar and co-axial bi-tori compared with a
straight-wire dimer for a gap distance $d=10$\,nm. Partial
contributions to the LDOS are shown as a function of wavelength for
a point in the center of the gap, both at fixed azimuthal number in
tori ($m=8$ for radius $b=750$\,nm and $m=4$ for $b=375$\,nm) and at
fixed parallel momentum in the straight dimer ($k_\parallel\approx
10.7\,\mu$m$^{-1}$, such that $k_\parallel=m/b$). The curves are
normalized to their maximum value.} \label{Fig4}
\end{figure}

We find it convenient to define a figure of merit $F$ for the
waveguides expressed as the ratio between the propagation distance
and the geometric mean of the mode diameter in the transverse
directions. The quantities $F^2$ and $F^3$ should be roughly
proportional to the number of logical elements that can be
integrated using a given waveguiding scheme with two and three
dimensional packing, respectively. We obtain $F\approx 540$ for the
wire dimer of Fig.\ \ref{Fig3} at a separation of 10\,nm. This has
to be compared with values of $F\lesssim 50$ for channel plasmon
polaritons \cite{BVD06} and particle arrays \cite{MKA03}. We
conclude that wire arrays yield high values of $F$, also improved
with respect to those obtained for single wires (e.g., $F\approx
100$ at 100\,nm radius and 1550\,nm wavelength). The decrease in
propagation length is the price to pay for plasmon confinement, but
wire arrays seem to perform optimally with respect to the figure of
merit $F$.


Reliable plasmon waveguides must be robust against fabrication
imperfections and sharp turns. Next, we show that gap waveguides
satisfy these requirements. In particular, curved waveguide paths
produce radiative losses originating in coupling to propagating
light waves when translational invariance is broken. We analyze this
effect in Fig.\ \ref{Fig4} both for non-identical co-planar tori and
for identical co-axial tori, using the prescription
$k_\parallel=m/b$ to compare with straight waveguide modes, where
$b$ is the toroidal radius (see insets) and $m$ is the azimuthal
momentum number. The calculations are performed using BEM,
specialized for axially-symmetric geometries \cite{Manja1}.
Radiative losses are still small compared to absorption for
$b=750$\,nm (cf. curves for straight wires and large-radius tori in
Fig.\ \ref{Fig4}, showing only $\sim 5$\% increase in peak width of
curved versus straight wires due to radiative losses in the former),
but they become sizable for shaper turns (the width increases by
40\% and 95\% for $b=375$\,nm in co-axial and co-planar torii,
respectively).

\begin{figure}[ht]
\includegraphics[width=70mm,angle=0]{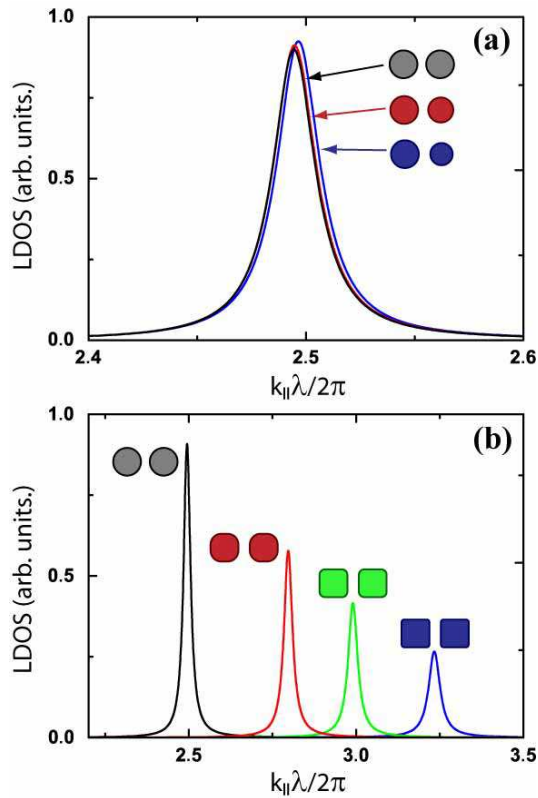}
\caption{Gap mode against variations of wire radius {\bf (a)} and
shape (from circular to square cross section) {\bf (b)}. The LDOS is
represented as a function of $k_\parallel$ for a point at the center
of each dimer and a wavelength $\lambda=1550$\,nm. One of the wires
in the dimers of (a) has a fixed radius of 100\,nm, while various
values of the radius are considered for the neighboring wire:
100\,nm, 90\,nm, and 80\,nm, from top to bottom. The distance
between wires is $d=10$\,nm in all cases. The horizontal diameter of
the wires in (b) is 200\,nm for all cross sections.} \label{Fig5}
\end{figure}

Guided gap plasmons are also robust against dimer asymmetries, as
shown in Fig.\ \ref{Fig5}(a) for fixed wavelength $\lambda=1550$\,nm
and gap distance $d=10$\,nm. Variations of up to 20\% in the
relative radius of neighboring wires produce just a small, tolerable
shift in $k_\parallel$. However, wire shape is a critical parameter,
which we study in Fig.\ \ref{Fig5}(b) through the transition from
circular to square cross section. This produces a shift of the gap
plasmon towards larger $k_\parallel$, consistent with the higher
degree of confinement that occurs when evolving from the line-like
contact of the circular wires to the planar waveguide defined by the
square wires, the guided modes of which have been the subject of
recent experimental investigation \cite{MK06}. This increase in
confinement is accompanied by peak broadening originating in larger
overlap of the gap mode with the metal (ohmic losses). The observed
extreme sensitivity to shape and separation of the wires imposes
severe limits to the precision required in the fabrication of the
arrays in order to maintain a homogenous mode wavelength along the
waveguide.

In conclusion, we have shown that gap plasmon modes existing in the
region defined by two neighboring nanowires are excellent candidates
to guide signals over tens of microns. These modes are quite robust
against both unintended variations of wire cross section and
curvature in short turns, and thus, gap plasmons can be guided with
minimum losses over complicated winding paths of micrometer
dimensions. Furthermore, gap modes are highly confined to the gap
region, so that inter-mixing between neighboring wire-dimers can be
minimized, thus preventing waveguide cross-talk and allowing
highly-integrated plasmonic circuits in three dimensional spaces.



\end{document}